\documentclass[preprint,12pt]{revtex4}

\usepackage[brazil, english]{babel}
\usepackage[utf8]{inputenc}
\usepackage{graphicx}
\usepackage{epsfig}
\usepackage{amssymb}
\usepackage{amsmath} 
\usepackage{slashed}
\usepackage{color}
\usepackage{hyperref}

\graphicspath{%
    {converted_graphics/}
    {/}
}
\begin{document}

\title{Comparison between holographic deformed AdS and soft wall models for fermions}
\author{Ayrton da Cruz Pereira do Nascimento}
\email[Eletronic address: ]{ayrton@pos.if.ufrj.br}
\author{Henrique Boschi-Filho}
\email[Eletronic address: ]{boschi@if.ufrj.br}  
\affiliation{Instituto de F\'{\i}sica, Universidade Federal do Rio de Janeiro, 21.941-909 - Rio de Janeiro - RJ - Brazil}

\begin{abstract} 
We compare the holographic dressed soft wall and the exponentially  deformed AdS models for spin 1/2 fermions. We present the dressed soft wall model and its analytical solutions for the left and right modes, and the corresponding spectra, also including modifications considering hyperfine spin-spin and meson cloud interactions, as well as anomalous dimensions. Then, we discuss the deformed AdS model for spin 1/2 fermions and present their effective Schr\"odinger equations for the left and right modes, for which only numerical solutions are available. Then, we consider a polynomial expansion of the effective potential of the deformed AdS model and show that in the quadratic  approximation it leads to exact analytical solutions comparable with the dressed soft wall model and obtain the corresponding spectra for left and right modes. We show a numerical comparison of the mass spectra of spin 1/2 baryons for the dressed soft wall and the deformed AdS models. We present a detailed relation between the quadratic approximation of the deformed AdS and the dressed soft wall models for their spectra, wave functions and comments on the deep inelastic scattering on both models. We find that these two models are {\sl not} equivalent even in the quadratic approximation, but it is possible to relate their left and right modes for particular choices of their  parameters. 
\end{abstract}


\maketitle

\tableofcontents


\section{Introduction}

The AdS/CFT correspondence, in its most well known form, relates a five dimensional gravitational theory in anti de-Sitter space with a conformal non-gravitational field theory in four dimensions
\cite{Maldacena:1997re, Aharony:1999ti}, in accordance with the holographic principle \cite{tHooft:1993dmi, Susskind:1994vu}. 
This relation is useful to describe difficult problems in non-perturbative models, since a weakly coupled 5-d gravitational theory is dual to a strongly coupled 4-d field theory. This has applications in many areas of physics and in particular for low energy QCD. This is made possible breaking the conformal symmetry of the original AdS/CFT proposal in different ways, which became known as the AdS/QCD models.

This was first realized in the holographic hard wall model, which constitutes a slice of finite size in the holographic coordinate in AdS space introduced  to describe the hard scattering \cite{Polchinski:2001tt} and spectra \cite{Boschi-Filho:2002xih, Boschi-Filho:2002wdj} of scalar glueballs, and deep inelastic scattering (DIS) of scalars and fermions \cite{Polchinski:2002jw}. Later, this model was extended to include the spectra of mesons \cite{Erlich:2005qh, deTeramond:2005su}, higher spin glueballs \cite{deTeramond:2005su, Boschi-Filho:2005xct} and baryons \cite{deTeramond:2005su}. Besides these achievements, one weak aspect of the hard wall model is that it leads to non-linear (parabolic) Regge trajectories, in contrast with the observed linear hadronic spectra. 

This point was overcome by the holographic soft wall model, which introduces an exponential factor in the AdS bulk action, leading to discrete spectra for mesons with linear Regge trajectories \cite{Karch:2006pv, Karch:2010eg, Colangelo:2008us}. 
Similar Regge trajectories are obtained from this model for glueballs \cite{Colangelo:2007pt}, although in this case the spectra is not good in comparison with lattice data. This discrepancy motivated the proposal of modifications in the original holographic soft wall model, as discussed for instance in \cite{Gursoy:2007cb, Gursoy:2007er, Li:2013oda}. 
In contrast with the hard wall, the original soft wall model does not give a discrete spectrum for fermions due to the bilinear form of the Dirac action. In this case the exponential factor is simply absorbed in a normalization of the fermionic fields. The first soft wall description of DIS was given by \cite{BallonBayona:2007qr} for mesons and where a hybrid model with an additional hard wall was used to discretize the fermionic spectra. 

An extension of the soft wall model which gives discrete spectra for spin 1/2 fermions was proposed in \cite{Abidin:2009hr}, where there is a supplementary interaction with the background field by using a scalar potential in the mass term. Since this term depends only on the holographic coordinate, this model is equivalent to a soft wall with a curvature dependent mass in the five dimensional AdS bulk which can also be understood as a dressed mass. This model has been successfully used to study mesons and baryons \cite{Branz:2010ub, Gutsche:2011vb}, and to obtain the DIS proton structure functions \cite{Braga:2011wa}. 

Recently, it appeared in the literature an AdS/QCD model with an exponential deformation in the AdS metric, instead of the case of the soft wall model where the exponential factor is included in the action. This exponentially deformed AdS model provides good spectra for hadrons of various  spins \cite{FolcoCapossoli:2019imm}, the DIS proton structure functions \cite{FolcoCapossoli:2020pks}, the pion  \cite{MartinContreras:2021yfz},  proton and neutron \cite{Contreras:2021epz} form factors, and in the Brownian motion 
\cite{Caldeira:2020sot, Caldeira:2020rir, Caldeira:2021izy}. 
This model was inspired by the obtention of a confining quark-antiquark potential from  the soft wall model \cite{Andreev:2006ct, Bruni:2018dqm}. It is important to notice that the deformed AdS model when applied to mesons gives exactly the same spectra as the original soft wall model. However, for any other hadrons the spectra are different.

Here, we consider two holographic AdS/QCD models, namely the soft wall model with a curvature dependent or dressed mass   \cite{Abidin:2009hr, Branz:2010ub, Gutsche:2011vb}, which for simplicity we are just calling it dressed soft wall and the exponentially deformed AdS/QCD  model  \cite{FolcoCapossoli:2019imm}. These models have been used recently to study proton structure functions in DIS \cite{Braga:2011wa, FolcoCapossoli:2020pks}, providing  similar results. One remarkable difference between these two models is that the dressed soft wall is analytically solvable for fermions while the in deformed AdS the corresponding equations are  only numerically tractable. We show a numerical comparison of the spectra of these models. In the case of the dressed soft wall model we also include hyperfine spin-spin and meson cloud interactions \cite{Gutsche:2011vb}, as well as anomalous dimensions affecting the fermionic spectra, following the ideas of Refs. \cite{Boschi-Filho:2012ijd, FolcoCapossoli:2016uns, Rodrigues:2016kez}. In both models, the spectra come from the solution of pairs of effective Schr\"odinger equations, usually labelled as left and right modes. In the case of the dressed soft wall, these equations reduce to a harmonic oscillators plus centrifugal barriers in opposition with the exponentially deformed AdS model with more involved effective potentials. 
Then, we consider a quadratic approximation for the effective potentials of the deformed AdS model find their analytical solutions and compare them with the ones coming from the dressed soft wall model. We show that the two models are {\sl not} equivalent even in the quadratic approximation, but it is possible to relate the left and right modes of both models for particular choices of their parameters. 

This work is organized as follows: In Section II, we briefly review the dressed soft wall model for spin 1/2 fermions, obtaining their effective Schr\"odinger like equations of motion, the corresponding analytical solutions, spectra,  and a modification of  the fermionic  spectra with the inclusion of corrections from hyperfine spin-spin and meson cloud interactions. In Subsection A, we present a fermionic spectrum modified by the inclusion of anomalous dimensions. 
In Section III, we discuss the exponentially deformed AdS model for spin 1/2 fermions and write their effective Schr\"odinger like equations of motion for the left and right modes, consider the expansion of the exponential deformation in the effective potentials as power series and restrict them to a quadratic approximation, which we solve analytically. Then, in Section IV, we present a brief discussion on the sign of the dilaton in these two models. In Section V, we compare spin 1/2 baryonic spectra in the dressed soft wall model (including hyperfine spin-spin and meson cloud interactions and anomalous dimensions) and the deformed AdS model. In section VI, we perform an analytical comparison of the dressed soft wall effective Schr\"odinger like equations and the corresponding ones for the deformed AdS in the quadratic regime and discuss the possible relations between these models in terms of spectra, wave functions and DIS interaction actions. Finally, in section VII, we present our conclusions. 
We also include three Appendices where we give some additional details of the material presented in the main text and discuss briefly the relation of the analytic soft walls with top-down D$p$-brane models \cite{Kirsch:2006he, Abt:2019tas, Nakas:2020hyo}.


\section{Dressed Soft Wall Model}\label{Dressed}

The original soft wall model is defined by the action \cite{Karch:2006pv, Colangelo:2008us}
\begin{equation}
    S=\int dz\,d^4{x}\sqrt{-g}\,e^{- \Phi(z)} \,\mathcal{L},\label{SWact}
\end{equation}
 where $\Phi(z) =\kappa^{2}\,z^{2}$ is the dilaton,   $\mathcal{L}$ is the field lagrangian and $g$ is the determinant of the metric $g_{mn}$ of the pure $AdS_{5}$ space: 
\begin{equation}
    ds^{2}=g_{mn}\,dx^{m}dx^{n}=\frac{\Omega^{2}}{z^{2}}(dz^{2}+\eta_{\mu\nu}dx^{\mu}dx^{\nu}),
\end{equation}
with $\Omega$ being the AdS radius, $z$ the holographic coordinate, and $\eta_{\mu\nu}$ the four dimensional flat Minkowski metric with signature $(-,+,+,+)$. The dilaton $\Phi(z)$ plays the role of a soft wall bringing in a discrete spectra, at least for mesons, and $\kappa$ which has dimension of mass is the IR scale of the model. 

A modified version of this model is required if we want to describe fermions with discrete mass spectra. This is so because in an standard fermionic action the exponential factor $e^{-\Phi(z)}$ would drop out from the equations of motion, thus not affecting their spectra, due to the bilinear form of the Dirac Lagrangian. This issue can be overcame introducing a potential ${\cal V}(z)$ in the mass term, which can be interpreted as a dressed mass in the action \cite{Abidin:2009hr}, such that the fermionic spectra becomes discrete. This is the dressed soft wall model. The fermionic action for spin 1/2 in this case is written as (following the notation of Ref. \cite{Braga:2011wa}):
\begin{equation}\label{FermionActionDSW}
    S=\int dz\,d^{4}x\,\sqrt{-g}\,e^{-\Phi(z)}\,\left[\frac{i}{2}\bar\Psi\,e^{m}_{a}\,\Gamma^{a}\,\mathcal{D}_{m}\Psi-\frac{i}{2}(\mathcal{D}_{m}\,\Psi)^{\dagger}\,\Gamma^{0}\,e^{m}_{a}\,\Gamma^{a}\,\Psi-\bar\Psi\left(\mu_5+{\cal V}(z)\right)\,\Psi\right],
\end{equation}
with $\mu_5$ a $5d$ constant fermion bulk mass and ${\cal V}(z)=\kappa^{2}\,z^{2}/\Omega$ was chosen to lead to exact solutions for the fermionic spectra. The dressed mass of this model is $\mu_5 + {\cal V}(z)$. The covariant derivative is given by 
\begin{equation}
    \mathcal{D}_{m}\equiv \partial_{m}+\frac{1}{2}\omega_{m}^{bc}\,\Sigma_{bc},
\end{equation}
where $\Sigma_{bc}=\frac{1}{4}[\Gamma_{b},\Gamma_{c}]$, with Dirac matrices $\Gamma^{a}=(\gamma^{\mu},-i\gamma^{5})$. The factor $e^{a}_{n}$ is the vielbein which couples the fermion to the  $AdS_{5}$ space, 
\begin{equation}
    e^{m}_{a}=\frac {z} {\Omega} \,\delta^{m}_{a},
\end{equation}
with $m=0, 1, 2, 3, 5$, and $\omega_{m}^{bc}$ is the spin connection, 
\begin{equation}
\omega^{a\,b}_{m}=- \frac {\Omega}{z} 
\left( \delta^a_z \delta^b_m - \delta^b_z \delta^a_m
\right). 
\end{equation}
After a rescaling of the fermionic field
\begin{equation}\label{Psi}
    \Psi(x^{\mu}, z)
    =  
    e^{\kappa^{2}\,z^{2}/2}\psi(x^{\mu}, z),
\end{equation}
the action reads
\begin{equation}
    S=\int dz\,d^{4}x\,\sqrt{-g}\,\left[\frac{i}{2}\bar\psi\,e^{m}_{a}\,\Gamma^{a}\,\mathcal{D}_{m}\psi-\frac{i}{2}(\mathcal{D}_{m}\,\psi)^{\dagger}\,\Gamma^{0}\,e^{m}_{a}\,\Gamma^{a}\,\psi-\bar\psi\left(\mu_5+\frac{\kappa^{2}z^{2}}{\Omega}\right)\,\psi\right].
\end{equation}

Now, considering a field decomposition into left and right chiral components
\begin{equation}
    \psi(x^{\mu}, z)=\psi_{_L}(x^{\mu}, z)+\psi_{_R}(x^{\mu}, z),
\end{equation}
one can write each chiral mode in terms of the four-dimensional Dirac spinors $ u_s (P)$ as
\begin{equation}
    \psi_{\frac{L}{R}}(x^{\mu}, z)= e^{i P\cdot x} \frac 1 2 (1 \mp \gamma^5) u_s (P) \frac{z^2}{\Omega^2}{f}_{\frac{L}{R}}(z), 
\end{equation}
where $P^\mu$ is the four-dimensional momenta of the hadron. The equations of motion can be written as \cite{Abidin:2009hr, Branz:2010ub, Gutsche:2011vb, Braga:2011wa} 
\begin{equation}
    \left[-\partial_{z}^{2}+\kappa^{4}\,z^{2}+2\,\kappa^{2}\left(m_5\mp\frac{1}{2} \right)+\frac{m_5(m_5\pm 1)}{z^{2}} \right]{f}^{n}_{\frac{L}{R}}(z)=M_{n}^{2}\,{f}^{n}_{\frac{L}{R}}(z),
\end{equation}
where 
\begin{equation}\label{m5}
    m_5=\mu_5\,\Omega
\end{equation}
is a dimensionless mass parameter of the five dimensional theory and $M_n$ are the hadron masses. The corresponding spectrum reads (see Appendix \ref{isotonic})
\begin{equation}
    M_{n\frac{L}{R}}^{2}=2\,\kappa^{2}(m_5\mp\frac{1}{2})
    + 2(2n+1)\kappa^2
    +\sqrt{\kappa^4[1+4m_5(m_5\pm 1)]}
    ;\qquad   (n=0, 1, 2, \dots),
\end{equation}
so that
\begin{equation}
    M_{n\frac{L}{R}}^{2}=2\kappa^{2}\left( m_5\mp \frac{1}{2} + 2n + 1 
    + \left|m_5 \pm \frac{1}{2}\right| \right)
    ;\qquad   (n=0, 1, 2, \dots).\label{MSWM||}
\end{equation}
If $m_5\ge 1/2$ the fractions with  $\mp$ and $\pm$ cancel, leading  exactly to the same spectrum for the right and left modes:
\begin{equation}
    M_{n}^{2}=4\,\kappa^{2}\left(n+m_5+\frac{1}{2}\right);\qquad \qquad  (n=0, 1, 2, \dots).\label{MSWM}
\end{equation}
On the other side, if $m_5 \le - 1/2$, the terms with $m_5$ cancel and the spectrum reads
\begin{equation}
    M_{n}^{2}=4\,\kappa^{2}\left(n+1\right);\qquad \qquad  (n=0, 1, 2, \dots).\label{MSWM12}
\end{equation}
If $m_5$ belongs to the interval $(-1/2,1/2)$, the spectrum for the left and right modes are different and given by Eq. \eqref{MSWM||}. 

In all of these cases, the corresponding Regge trajectories are linear as expected for the soft wall, and the normalized solutions are given by
\begin{align}
    {f}^{n}_{_{L/R}}(z)&=\sqrt{\frac{2\,\Gamma(n+1)}{\Gamma(n+|m_5\pm 1/2|+1)}}\, |\kappa|^{|m_5\pm 1/2|+1}
    \,z^{|m_5\pm 1/2|+1/2}\,
    e^{-\kappa^{2}z^{2}/2}\,
    L^{|m_5\pm 1/2|}_{n}(\kappa^{2}z^{2}),  
\end{align}
where $L^{p}_{n}(\kappa^2\,z^{2})$ are the associated Laguerre polynomials (see Appendix \ref{isotonic}).

In particular, for $m_5\ge 1/2$, the normalized solutions  ${f}^{n}_{L/R}(z)$ are given by 
\begin{subequations}
\begin{align}
    {f}^{n}_{_L}(z)&=\sqrt{\frac{2\,\Gamma(n+1)}{\Gamma(n+m_5+3/2)}}\,|\kappa|^{m_5+3/2}\,z^{m_5+1}\,e^{-\kappa^{2}z^{2}/2}\,L^{m_5+1/2}_{n}(\kappa^{2}z^{2}),\\
    {f}^{n}_{_R}(z)&=\sqrt{\frac{2\,\Gamma(n+1)}{\Gamma(n+m_5+1/2)}}\,|\kappa|^{m_5+1/2}\,z^{m_5}\,e^{-\kappa^{2}z^{2}/2}\,L^{m_5-1/2}_{n}(\kappa^{2}z^{2}), 
\end{align}
\end{subequations}
recovering the results of Ref. \cite{Braga:2011wa}.

In order to find a baryonic spectra from the above discussion, one can start from the standard relation of the AdS/CFT dictionary for spin 1/2 fermions,  $|m_5|=\Delta - 2$, and consider that each fermion in $d=4$ spacetime dimensions has conformal dimension 3/2, so a baryon with three constituent quarks imply $\Delta=9/2$, so that
$m_5=\pm 5/2$. Taking for simplicity $m_5= 5/2$, then Eq. \eqref{MSWM} reads
\begin{equation}
    M_{n}^{2}=4\,\kappa^{2}
    \left(n+3\right);\qquad \qquad  (n= 0, 1, 2,  \dots).\label{MSWMa}
\end{equation}
In Ref. \cite{Gutsche:2011vb}, the authors considered the mass parameter for spin 1/2 fermions in $d=4$ spacetime dimensions as  $m_5=\ell+3/2$, where $\ell=0,1$. Then, Eq. \eqref{MSWM} gets 
\begin{equation}
    M_{n,\ell}^{2}=4\,\kappa^{2}\left(n+\ell+2\right);\qquad \qquad  (n=0, 1, 2, \dots).\label{MSWML}
\end{equation}
and the wave functions can be rewritten as 
\begin{subequations}
\begin{align}
    {f}^{n}_{_L}(z)&=\sqrt{\frac{2\,\Gamma(n+1)}{\Gamma(n+\ell+3)}}\,\kappa^{\ell+3}\,z^{\ell+5/2}\,e^{-\kappa^{2}z^{2}/2}\,L^{\ell+2}_{n}(\kappa^{2}z^{2}),\\
    {f}^{n}_{_R}(z)&=\sqrt{\frac{2\,\Gamma(n+1)}{\Gamma(n+\ell+2)}}\,\kappa^{\ell+2}\,z^{\ell+3/2}\,e^{-\kappa^{2}z^{2}/2}\,L^{\ell+1}_{n}(\kappa^{2}z^{2}).
\end{align}
\end{subequations}
In this case, the spectra and eigenfunctions are degenerate since $n=0,1,2,...$ and $\ell=0,1$. So, using some phenomenological arguments related to hyperfine  spin-spin and meson cloud  interactions, they find a formula for baryonic spectra \cite{Gutsche:2011vb} 
\begin{equation}
    M_{_B}^{2}=4\,\kappa^{2}
    \left[n+\ell+2 +\alpha_{_{\rm HF}}(S-1)-\beta_{_{\rm MC}}
    \right];\qquad \qquad  (n= 0, 1, 2,  \dots).\label{Gutsche_et_al}
\end{equation}
with $\kappa=$ 500 MeV, $\alpha_{_{\rm HF}}=0.636$ and $\beta_{_{\rm MC}}=0.800$. With these parameters they were able the obtain, for instance, the spectra of $N(1/2^+)$ baryons with $S=1/2$ and $\ell=0$, which are shown in Table \ref{t1}. 


\subsection{Anomalous Dressed Soft Wall Model}

Alternatively, it is also possible to modify the spectrum given, for instance, by Eqs. \eqref{MSWM} and \eqref{MSWMa}, taking into account anomalous dimensions in the conformal dimension of the baryonic operators. This follows the idea of introducing anomalous dimensions in the soft wall model for glueball operators \cite{Boschi-Filho:2012ijd, FolcoCapossoli:2016uns, Rodrigues:2016kez}, and was also used in the numerical solution of the deformed AdS model for baryons \cite{FolcoCapossoli:2020pks}. 

Here, we use the idea of introducing an anomalous dimension in the analytical baryonic spectrum, given by Eq. \eqref{MSWMa}. 
Then, taking  $|m_5|=\Delta_{\rm canonical} - 2 + \gamma$, one has 
\begin{equation}
    M_{{n}_{A}}^{2}=4\,\kappa^{2}
    \left(n+3+\gamma\right);\qquad \qquad  (n= 0, 1, 2,  \dots), \label{MSWM2}
\end{equation}
where the anomalous dimension $\gamma$ is usually a function of the spin. For instance, if one has  $\gamma=-2$, one finds 
\begin{equation}
    M_{{n}_{A}}^{2}=4\,\kappa^{2}
    \left(n+1\right);\qquad \qquad  (n= 0, 1, 2,  \dots).\label{MSWMA}
\end{equation}
Clearly, comparing eqs. \eqref{Gutsche_et_al}, \eqref{MSWM2},  and \eqref{MSWMA}, one sees that the consideration of hyperfine spin-spin and meson cloud interactions in Ref. \cite{Gutsche:2011vb} is equivalent to including anomalous dimensions in the conformal dimension of baryonic operators. The masses calculated according to the above formula are also presented in Table \ref{t1}.


\section{Deformed AdS Model}
\label{Deformed}

In this section we discuss the deformed AdS model. The 5-dimensional action is given by \cite{FolcoCapossoli:2019imm}
\begin{equation}
    \tilde S=\int d\tilde z\,d^{4}x\,\sqrt{-\tilde g}\,\tilde{\mathcal{L}},
\end{equation}
where $\tilde{\mathcal{L}}$ is the lagrangian density and $\tilde g$ is the determinant of the metric of the exponentially deformed $AdS_{5}$ space 
\begin{equation}\label{deform_metric}
    d\tilde s^{2}=\tilde g_{mn}\,d\tilde x^{m}d\tilde x^{n}=e^{2\,A(\tilde z)}\,(d\tilde z^{2}+\eta_{\mu\nu}dx^{\mu}dx^{\nu}),
\end{equation}
 $\tilde z$ being the holographic coordinate and
\begin{equation}
    A(\tilde z)=-\log\left(\frac{\tilde z}{\tilde\Omega}\right)+\frac{1}{2}\,k\,\tilde z^{2}.\label{warpf}
\end{equation}
 The constant $k$ has dimension of mass squared. Note that in the asymptotic limit $k\rightarrow 0$, the pure AdS space with radius $\tilde\Omega$ is obtained. 

Now, we address the question of how to describe fermionic fields in this deformed background model. 
 To do so,  we follow the steps in \cite{FolcoCapossoli:2020pks}, 
starting with the fermionic action given by
\begin{equation}
    \tilde S=\int d\tilde z \, d^{4}x \,\sqrt{\tilde g}\,\tilde{\bar\Psi}(\tilde{\slashed{D}}-\tilde \mu_{5})\tilde\Psi,\label{Action}
\end{equation}
where $\tilde \mu_{5}$ is a constant fermionic bulk mass and the operator $\tilde{\slashed{D}}$ is defined as:  
\begin{equation}
    \tilde{\slashed{D}}\equiv \tilde g^{mn}\,\tilde e^{a}_{n}\,\Gamma_{a}\left(\partial_{m}+\frac{1}{2}\tilde\omega_{m}^{bc}\,\Sigma_{bc}\right). 
\end{equation}
Note that the Dirac matrices $\Gamma_{a}$ and $\Sigma_{bc}$ were defined in the previous section, and here the vielbein $\tilde e^{a}_{n}$ is given by
\begin{equation}
    \tilde e^{a}_{m}=e^{A(\tilde z)}\,\delta^{a}_{m}. 
\end{equation}
The spin connection $\tilde \omega_{m}^{bc}$ is 
\begin{equation}
    \tilde\omega^{a\,b}_{m}=\tilde e^{a}_{n}\,\partial_{m}\tilde e^{n\,b}+\tilde e^{a}_{n}\,\tilde e^{p\,b}\,\tilde\Gamma^{n}_{p\,m}, 
\end{equation}
with $m=0, 1, 2, 3, 5$, and $\tilde\Gamma^{n}_{p\,m}$ represent the Christoffel symbols which can be written as
\begin{equation}
    \tilde\Gamma^{p}_{m\,n}=\frac{1}{2}\,\tilde g^{p\,q}(\partial_{n}\,\tilde g_{m\,q}+\partial_{m}\,\tilde g_{n\,q}-\partial_{q}\,\tilde g_{m\,n}). 
\end{equation}

From Eq. (\ref{Action}), one can derive the following equation of motion \cite{FolcoCapossoli:2020pks} 
\begin{equation}
    (e^{-A(\tilde z)}\,\gamma^{5}\partial_{\tilde z}+e^{-A(\tilde z)}\, \gamma^{\mu}\partial_{\mu}+2\, A^{\prime}(\tilde z)\gamma^{5}-\tilde \mu_{5})\,\tilde\Psi=0,
\end{equation}
and prime denote derivative with respect to $\tilde z$. As before, the fermionic fields are written in terms of left and right chiral components
\begin{equation}
    \tilde\Psi(x^{\mu}, \tilde z) =\tilde\Psi_{_L}(x^{\mu}, \tilde z) +\tilde\Psi_{_R}(x^{\mu}, \tilde z), \label{Pcompts}
\end{equation}
which can be written as
\begin{equation}
    \tilde\Psi_{\frac{L}{R}}(x^{\mu},\tilde z)= e^{i P\cdot x} \frac 1 2 (1 \mp \gamma^5) u_s (P) \frac{\tilde z^2}{{\tilde\Omega}^2}\tilde{f}_{\frac{L}{R}}(\tilde z)\,e^{-2\,A(\tilde z)}. 
\end{equation}
Then, one gets an effective Schrödinger equation for each chiral mode $\tilde f_{\frac{L}{R}}(\tilde z)$ \cite{FolcoCapossoli:2020pks}
\begin{equation}
    [-\partial^{2}_{\tilde z}+V_{\frac{L}{R}}(\tilde z)]\,\tilde f_{\frac{L}{R}}^{n}(\tilde z)=\tilde M_{n}^{2}\,\tilde f^{n}_{\frac{L}{R}}(\tilde z),\label{SchEq}
\end{equation}
where $\tilde M_{n}$ is the $n$-th fermionic mass in four dimensions  in the exponentially deformed AdS model and $V_{\frac{L}{R}}(\tilde z)$ are the chiral potentials 
\begin{equation}
    V_{\frac{L}{R}}(\tilde z)=\tilde \mu_{5}^{2}\, e^{2\,A(\tilde z)}\mp \, e^{A(\tilde z)}\,\tilde \mu_{5}\,A^{\prime}(\tilde z). \label{Potential}
\end{equation}
With these potentials, the effective Schrödinger equations,  Eq. (\ref{SchEq}), in general, cannot be solved analytically and were solved numerically in \cite{FolcoCapossoli:2019imm, FolcoCapossoli:2020pks}. In the next subsection, we implement a perturbative approach for these potentials, leading to approximate analytical solutions for the chiral modes $\tilde f^{n}_{\frac{L}{R}}(\tilde z)$.


\subsection{Quadratic approximation}

In order to perform an analytical comparison of the two models, here we make a quadratic approximation on the  exponentially deformed AdS model. In the deformed AdS model, we have effective Schrödinger equations for each chiral mode $\tilde f^{n}_{R/L}(\tilde z)$ with chiral potentials, Eq.  \eqref{Potential}. These potentials depend on exponentials of the warp factor $A(\tilde z)$, given by equation \eqref{warpf}, which can be written as
\begin{align}
    e^{A(\tilde z)}&=\frac{\tilde\Omega}{\tilde z}\,e^{k\,\tilde z^{2}/2}.
\end{align}

Introducing  
$ \xi=k\,\tilde z^{2}$, 
%
%
%
one can expand the exponentials to second order in this new variable, to obtain
\begin{subequations}
\begin{align}
    e^{A(\tilde z)}&=\frac{\tilde\Omega}{\tilde z} \,e^{\xi/2}=\frac{\tilde\Omega}{\tilde z }\left(1+\frac{\xi}{2}+\frac{1}{8}\xi^{2}+ {\cal O} (\xi^3) \right)\approx \frac{\tilde\Omega}{z}+\frac{k\,\tilde z\,\tilde\Omega}{2}+\frac{1}{8}k^{2}\,\tilde z^{3}\, \tilde\Omega ,\label{Aofz}\\
     e^{2\,A(\tilde z)} &=\frac{\tilde\Omega^{2}}{\tilde z^{2}}\,e^{\xi}= \frac{\tilde\Omega^{2}}{\tilde z^{2}}\, \left(1+\xi+\frac{1}{2}\,\xi^{2}+ {\cal O} (\xi^3)\right)\approx\frac{\tilde\Omega^{2}}{\tilde z^{2}} +\tilde\Omega^{2}\,k+\frac{\tilde\Omega^{2}}{2}\,k^{2}\,\tilde z^{2}.\label{2A}
\end{align}
\end{subequations}

We also have
\begin{equation}
    A^{\prime}(\tilde z)=\frac{d}{d\,\tilde z}\left(-\log\left(\frac{\tilde z}{\tilde\Omega}\right)+\frac{k\,\tilde z^{2}}{2}\right)=-\frac{1}{\tilde z}+k\,\tilde z.\label{Aprime}
\end{equation}
Using (\ref{Aofz}) alongside with (\ref{Aprime}), and keeping terms up to $\tilde z^2$, we get
\begin{equation}
    e^{A(\tilde z)}\,A^{\prime}(\tilde z) =\left(\frac{\tilde\Omega}{\tilde z}+\frac{k\,\tilde z\, \tilde\Omega}{2}+\frac{1}{8}k^{2}\,\tilde z^{3}\,\tilde\Omega\right)\left(-\frac{1}{\tilde z}+k\,\tilde z\right) \approx -\frac{\tilde\Omega}{\tilde z^{2}}+\frac{k\,\tilde\Omega}{2}+\frac{3k^{2}\,\tilde z^{2}\,\tilde\Omega}{8}.\label{etimesAprime}
\end{equation}
Substituting (\ref{2A}) and (\ref{etimesAprime}) in (\ref{Potential}), we obtain the effective potential 
\begin{equation}
    V_{\frac{L}{R}}(\tilde z)=\tilde \mu_{5}^{2}\tilde\Omega^2\left(\frac{1}{\tilde z^{2}}+k+\frac{1}{2}\,k^{2}\,\tilde z^{2}\right)\mp \tilde \mu_{5}\tilde\Omega\left(-\frac{1}{\tilde z^{2}}+\frac{k}{2}+\frac{3}{8}\,k^{2}\,\tilde z^{2}\right),
\end{equation}
or equivalently 
\begin{equation}
    V_{\frac{L}{R}}(\tilde z)=\frac{k_{1\frac{L}{R}}}{\tilde z^{2}}+k_{2\frac{L}{R}}+{k_{3\frac{L}{R}}}\,\tilde z^{2},
\end{equation}
with
\begin{subequations}
\begin{align}
k_{1\frac{L}{R}}&=\Tilde m_{5}^{2}\pm \Tilde m_{5},\\
k_{2\frac{L}{R}}&=\left(\Tilde m_{5}^{2}\mp \frac{\Tilde m_{5}}{2}\right)\,k,\\
k_{3\frac{L}{R}}&=\left(\frac{\Tilde m_{5}^{2}}{2}\mp \frac{3\,\Tilde m_{5}}{8}\right)\,k^{2},
\end{align}
\end{subequations}
where we have introduced the dimensionless parameter
\begin{equation}\label{m5tilde}
\Tilde m_{5}=\tilde \mu_{5}\,\tilde\Omega.
\end{equation}
Then, the effective Schr\"odinger equations (\ref{SchEq}) for each chiral mode in the quadratic approximation read
\begin{equation}
    \left[-\partial_{\tilde z}^{2}+\frac{k_{1\frac{L}{R}}}{\tilde z^{2}}+k_{2\frac{L}{R}}+k_{3\frac{L}{R}}\,\tilde z^{2}\right]\,\tilde f_{\frac{L}{R}}^{n}(\tilde z)=M_{n}^{2}\,\tilde f_{\frac{L}{R}}^{n}(\tilde z).\label{eqS}
\end{equation}
We thus have a Schrödinger equation with a harmonic term plus a centrifugal/centripetal barrier term falling off as $1/\tilde z^{2}$. Such equations have analytical solutions which can be read off as (see Appendix \ref{isotonic}) 
\begin{equation}
    \tilde f_{\frac{L}{R}}^{n}(\tilde z) =\sqrt{\frac{2n!\,\left(k_{3\frac{L}{R}}\right)^{\frac{1}{2}\left(1+\phi_{\frac{L}{R}}\right)}}{\Gamma(n+\phi_{\frac{L}{R}}+1)}}\, \tilde z^{\left(\frac{1}{2}+\phi_{\frac{L}{R}}\right)}\,\exp{\left(-\frac{1}{2}\,\sqrt{k_{3\frac{L}{R}}}\,\tilde z^{2}\right)}\,L^{\left(\phi_{\frac{L}{R}}\right)}_{n}\left(\sqrt{k_{3\frac{L}{R}}}\,\tilde z^{2}\right),\label{SolLa}
\end{equation}
with
\begin{equation}
    \phi_{\frac{L}{R}}=\frac{1}{2}\sqrt{1+4\,k_{1\frac{L}{R}}}.
\end{equation}
The spectrum is given by 
\begin{align}\label{DefEspQuad}
    M_{n\frac{L}{R}}^{2}
    &=k_{2\,\frac{L}{R}}+2\left[(2\,n+1)\,\sqrt{k_{3\,\frac{L}{R}}}\right]+\sqrt{k_{3\,\frac{L}{R}}\,(1+4\,k_{1\,\frac{L}{R}})}\nonumber\\
        &=
        (\Tilde m_{5}^{2}\mp \frac{\Tilde m_{5}}{2})\,k+2\,|k|\,\sqrt{\left(\frac{\Tilde m_{5}^{2}}{2}\mp \frac{3\,\Tilde m_{5}}{8}\right)}\,\left(2n+1+ \left|\tilde m_5 \pm \frac12\right|\right)
\end{align}
with $n=0, 1, 2,\dots$. If $\tilde m_5 \ge 1/2$, then the spectrum reduces to 
\begin{subequations}
\begin{align}\label{DefEspQuad+}
    M_{n\frac{L}{R}}^{2>}
        &=
        (\Tilde m_{5}^{2}\mp \frac{\Tilde m_{5}}{2})\,k+2\,|k|\,\sqrt{\left(\frac{\Tilde m_{5}^{2}}{2}\mp \frac{3\,\Tilde m_{5}}{8}\right)}\,\left(2n+1+ \tilde m_5 \pm \frac12\right)
\end{align}
and if $\tilde m_5 \le -1/2$, it gives 
\begin{align}\label{DefEspQuad-}
    M_{n\frac{L}{R}}^{2<}
        &=
        (\Tilde m_{5}^{2}\mp \frac{\Tilde m_{5}}{2})\,k+2\,|k|\,\sqrt{\left(\frac{\Tilde m_{5}^{2}}{2}\mp\frac{3\,\Tilde m_{5}}{8}\right)}\,\left(2n+1- \tilde m_5 \mp \frac12\right).
\end{align}
\end{subequations}
In the case where $\tilde m_5$ belongs to the interval $(-1/2, 1/2)$, the spectrum is given by Eq. \eqref{DefEspQuad}. So, we see that, in general, the masses for the  $L$ and $R$ sectors in the deformed AdS model, in this approximation, are different. In Section V, we are going to see how the masses in the  $L$ and $R$ sectors can be related.


\section{Comment on the Signs of the Dilaton for Fermions}\label{Sign}

Initially, in this section, before dealing with the question of the signs of the dilaton for fermions in soft wall models, for clarity, we briefly review the case for vector mesons. 

The soft wall was proposed in Ref. \cite{Karch:2006pv} for vector mesons, where the question of the sign of the dilation was touched and further discussed in Ref. \cite{Karch:2010eg}, rewriting the action as 
\begin{equation}
    S=\int dz\,d^4{x}\sqrt{-g}\,e^{- a z^2} \,\mathcal{L},\label{SWa}
\end{equation}
where it as shown that the case $a<0$ implies a spurious massless mode in the vector meson spectra, in contrast with the more interesting case $a>0$ where there are only massive physical modes. 

On the other side, in order to obtain a confining quark-antiquark potential, the authors of Ref. 
\cite{Andreev:2006ct} introduced an exponential in the AdS metric as
\begin{equation}\label{AZ}
    d s^{2}=\frac{\Omega^2 }{z^2}\, h_2(z) \,(dz^{2}
+dx^{i}dx^{i}); \qquad h_2(z)=e^{\frac 12 c_2 z^2}, 
\end{equation}
where the constant $c_2>0$. This result can generalized to a metric of the form \cite{Bruni:2018dqm}
\begin{equation}
    d s^{2}=\frac{\Omega^2 }{z^2} \, h_n(z)\,(dz^{2}
+dx^{i}dx^{i}); \qquad h_n(z)=e^{\frac 1n (c_n z)^n}, 
\end{equation}
with the restriction $n>0$ besides $c_n >0$, so that a linear confining quark-antiquark potential is also obtained in this case.   

It is interesting to observe that an alternative soft wall model for {\sl vector mesons}  starting with metric \eqref{AZ} and no dilaton in the action is equivalent to the original soft wall model \eqref{SWa} with $a<0$, which is problematic in the vector meson spectra. 

Now, we comment on the signs of the dilaton for the fermionic models discussed in Sections \ref{Dressed} and \ref{Deformed} above. 

First of all, it is important to remark that the sign of the dilaton in the original and dressed soft wall models for {\sl fermions} (Section \ref{Dressed}) is {\sl irrelevant}, since it does not modify the spectra or any other of their properties. This is so, because the exponential of the dilaton field in the fermionic action can be factored out of the corresponding equations of motion. For this reason, a potential similar to the dilaton field is added in the dressed model to the bare fermionic bulk mass, so that these particles acquire a discrete mass spectrum. 

On the other hand, in Section \ref{Deformed}, the positive sign of quadratic exponential deformation of the metric, Eqs. \eqref{deform_metric} and \eqref{warpf}, comes from a numerical fit of this model for baryons of spin 1/2, 3/2 and 5/2 to describe conveniently their spectra. In fact, this deformed AdS model when applied to glueballs of even and odd spins also have a positive sign for the quadratic exponential deformation of the metric, as in the case of fermions. The exceptions are the cases of scalar and vector mesons for which the deformed AdS model is better described by a quadratic exponential deformation with negative sign \cite{FolcoCapossoli:2019imm}. This is why in Section \ref{Deformed}, we used the quadratic exponential deformation with a positive sign. 

In view of these observations, we see that the constants $\kappa$, Eq. \eqref{FermionActionDSW}, and $\sqrt{k}$, Eq. \eqref{warpf}, play the role of IR scales in the dressed soft wall  and the deformed AdS models for fermions, in Sections \ref{Dressed} and \ref{Deformed}, respectively. Actually, in Section 
\ref{Numerical} below, we present a numerical comparison of the baryonic spectra obtained from these models where the values of these constants are of the same order, in the range 200-500 MeV, and in Section \ref{Analytic}, we are going to relate these IR scales of both models when the effective potential of the deformed AdS model is quadratically approximated.


\section{Numerical comparison of the models}\label{Numerical}

It is interesting to calculate masses within these models for spin 1/2 hadrons, to compare the outputs. Then, taking the case of  $N(1/2^+)$ baryons the masses were calculated using the dressed soft wall, corresponding to Eqs. \eqref{MSWMa}, \eqref{Gutsche_et_al}, and \eqref{MSWMA}, with  
$\kappa=$~271~MeV,   
$\kappa=$~500~MeV, and 
$\kappa=$~470~MeV, 
respectively, using  939~MeV as inputs. The   masses  for the deformed AdS model, were obtained solving numerically Eq.\eqref{SchEq} with $k= 205^2$\,{\rm MeV}$^2$. These masses are presented in Table \ref{t1} together with experimental data coming from PDG \cite{Workman:2022ynf}.

\begin{table}[h]
\centering
\centering
\begin{tabular}{|c|c|c|c|c|c|c|}
\hline
 &  \multicolumn{6}{c|}{Baryons $N (1/2^{+}$)}  \\  
\cline{2-7}
 & $N$ baryon & \,\, $M_{\rm exp}$ \, \, & Eq.\eqref{MSWMa} & Eq.\eqref{Gutsche_et_al} & Eq.\eqref{MSWMA}  & Eq.\eqref{SchEq}   \\
\hline \hline
 \, $n=0$  \,                                    
&\, $N(939)$\, & 939 & 939 &  \, 939  \, & 939 &  \, 987 \, \\ \hline
$n=1$&\, $N(1440)$\, & 1370 & \, 1085  \,&1372 & 1330 &\, 1264  \, \\ \hline
$n=2$&\, $N(1710)$\, & 1700 & \, 1213  \,&1698 & 1630 & \, 1531 \, \\ \hline
$n=3$&\, $N(1880)$\, & 1860 &\, 1329  \, &1970 & 1880 &\, 1791 \,  \\ \hline
$n=4$&\, $N(2100)$\, & 2100 & \, 1435  \, & 2209 & 2100 &\, 2046  \, \\ \hline
$n=5$&\, $N(2300)$\, & 2300 & 1534  & 2425 &\, 2300 \,&  2296  \\ \hline
\end{tabular}
\caption{Masses of $N(1/2^+)$ baryons expressed in MeV. The column $n=0,\, 1,\, 2,\,  \dots$ represents the holographic radial quantum number excitation.  The column $M_{\rm exp}$ represents the experimental data coming from PDG \cite{Workman:2022ynf}. The columns Eqs. \eqref{MSWMa}, \eqref{Gutsche_et_al}, and \eqref{MSWMA}, give the spectra corresponding to these equations in the dressed soft wall, discussed in Section II, 
with  $\kappa=$~271~MeV,  $\kappa=$~500~MeV, and $\kappa=$~470~MeV, respectively, using  939 MeV as inputs. The column Eq.\eqref{SchEq} represents the masses  obtained numerically within the deformed AdS model, Section III, with $k= 205^2$\,{\rm MeV}$^2$.}
\label{t1}
\end{table}

One can note from Table \ref{t1} that the masses obtained for the dressed soft wall model, starting from 939 MeV, from Eqs. \eqref{MSWMa}, \eqref{Gutsche_et_al}, and \eqref{MSWMA}, vary significantly, depending on additional modeling as the inclusion of 
hyperfine spin-spin and meson cloud interactions, Eq. \eqref{Gutsche_et_al}, or anomalous dimensions, Eq. \eqref{MSWMA}. In particular, one sees from the column of Table \ref{t1}, corresponding to Eq. \eqref{MSWMa}, that these masses are between 20 and 33$\%$ lower than the experimental data from PDG. After including hyperfine spin-spin and meson cloud interactions, as in Ref. \cite{Gutsche:2011vb}, one obtains from Eq. \eqref{Gutsche_et_al}, very good results, specially for $n=1,2$, with difference with PDG lower than 0.15$\%$, while for $n=3,4,5$ the masses are higher than that of PDG by 5 to 6$\%$. Still in the dressed soft wall model, this time with anomalous dimensions, the masses given by Eq. \eqref{MSWMA}. For $n=1,2$ they are lower than that of PDG by 3 to 4$\%$, while for $n=3,4,5$ they are very good compared with experimental data. Finally, from the deformed AdS model, the masses coming from the numerical solution to Eq. \eqref{SchEq}, are all lower than that of PDG, for $n=1,2$, with deviation of 8 and 10$\%$, respectively, while  for $n=3,4,5$, with 3.7, 2.6 and 0.2$\%$, respectively. Note that in this case, the mass for the nucleon ground state is 5.1$\%$ higher than the well known value of 939 MeV. 


%
\section{Analytic comparison of the models}\label{Analytic}
\label{Analytic}

%
The models we described above have different approaches to spin 1/2 fields in the $AdS_{5}$ or asymptotically $AdS_{5}$ spaces. The deformed background space model presents a warp factor in the metric, with no extra factor in the action. Conversely, the dressed soft wall model presents no deformations in the metric, but the addition of a $z^2$ dependent factor in the mass term of the action, besides the dilaton field.

In particular, when applied to fermions we see that they lead to distinct equations of motion. In the dressed soft wall model we have
\begin{equation}
    \left[-\partial_{z}^{2}+\kappa^{4}\,z^{2}+2\,\kappa^{2}\left(m_5\mp\frac{1}{2} \right)+\frac{m_5(m_5\pm 1)}{z^{2}} \right] {f}^{n}_{\frac{L}{R}}(z)=M_{n}^{2}\, {f}^{n}_{\frac{L}{R}}(z),\label{SW}
\end{equation}
while in the deformed AdS model we get
\begin{equation}
    \left\{-\partial^{2}_{\tilde z}+\tilde m_{5}\, e^{\,A(\tilde z)}\left[\tilde m_{5}\, e^{\,A(\tilde z)}\mp \,A^{\prime}(\tilde z)\right]\right\}\,\psi_{\frac{L}{R}}^{n}(\tilde z) =M_{n}^{2}\,\psi^{n}_{\frac{L}{R}}(\tilde z),\label{DADS}
\end{equation}
which, in the quadratic approximation, reads
\begin{equation}
    \left[-\partial^{2}_{\tilde z}+\Tilde m_{5}\left(\frac{\Tilde m_{5}}{2}\mp\frac{3}{8}\right)\,k^{2}\,\tilde z^{2}+\Tilde m_{5}\left(\Tilde m_{5}\mp\frac{1}{2}\right)\,k+\frac{\Tilde m_{5}\,(\Tilde m_{5}\pm 1)}{\tilde z^{2}}\right]\, \psi_{\frac{L}{R}}^{n}(\tilde z)=M_{n}^{2}\,\psi_{\frac{L}{R}}^{n}(\tilde z).\label{AppEq}
\end{equation}
Comparing Eqs. (\ref{SW}) and \eqref{AppEq}, we see that they have the same structure, that is, a harmonic and a centrifugal/centripetal barrier term plus a constant. Both have similar solutions in terms of the associated Laguerre polynomials (see Appendix \ref{isotonic}). The equivalence of the left and right modes of  Eqs.  (\ref{SW}) and (\ref{AppEq}) hold, if and and only if the following conditions are satisfied: 
\begin{subequations}
\begin{align}
    m_5\,(m_5\pm 1)&=\Tilde m_{5}\,(\Tilde m_{5}\pm 1),\label{1}\\
    2\,\kappa^{2}\,\left(m_5\mp\frac{1}{2}\right)&=\Tilde m_{5}\,\left(\Tilde m_{5}\mp\frac{1}{2}\right)\,k,\label{2}\\
    \kappa^{4}&=\Tilde m_{5}\,\left(\frac{\Tilde m_{5}}{2}\mp \frac{3}{8}\right)\,k^{2}.\label{3}
\end{align}
\end{subequations}
Let us now look at what these conditions imply. 

First of all, equation (\ref{1}) imply one of the following conditions: 
\begin{subequations}
\begin{align}
 (i) & \qquad\qquad  m_5=\Tilde m_{5},\\
 (ii) & \qquad\qquad  m_5=-\Tilde m_{5}\mp 1.
\end{align} 
\end{subequations}
We now study the the consistency of these solutions with the other two conditions, Eqs. \eqref{2} and \eqref{3}. 

Let us discuss here the moist interesting case $(i)$ where  $m_5=\Tilde m_{5}$, and leave the other to  Appendix \ref{case(ii)}. 
Imposing the condition $m_5=\Tilde m_{5}$ into Eq. (\ref{2}), one gets
\begin{eqnarray}
2\,\kappa^{2}=\Tilde m_{5}\,k.\label{kappaK}
\end{eqnarray}
Then, substituting this equation into Eq. (\ref{3}), one finds the relations 
\begin{eqnarray}
m_{5\pm} = \Tilde m_{5\pm} = \pm \frac{3}{2}, \qquad \qquad {\rm and} \qquad \qquad 
    \,k_{\pm}\,= \pm \frac{4}{3}\kappa^{2},  \label{m5metilde3/2}
\end{eqnarray}
where the plus (minus) signs refer to the left (right) modes and imply relations between the mass parameters and scales of the dressed soft wall and deformed AdS models. 

Since $\kappa$ is real, one concludes that the conditions, Eqs. \eqref{1}, \eqref{2}, and \eqref{3}, imply that the IR scales $k_{\pm}$ of the quadratic approximation for the deformed AdS model must have different signs for its left $(+)$ and right $(-)$ modes. Also, the left modes should have mass parameters $m_{5+}=\Tilde m_{5+}=+3/2$, while for the right modes $m_{5-}=\Tilde m_{5-}=- 3/2$, in both models. 

These differences between the left and right modes of both models suggest that they are not equivalent even in the quadratic approximation considered for the deformed AdS case. This non-equivalence of the models will be confirmed by the discussion of the spectra below.


\subsection{Spectra}

Now, to compare the spectra of these analytical models, consider the effective Schr\"odinger equation for the dressed soft wall, Eq. (\ref{SW}). 
Using $m_{5\pm}=\Tilde m_{5\pm}=\pm 3/2$ and $\,k_{\pm}\,= \pm \frac{4}{3}\kappa^{2}$, one finds the equations 
\begin{equation}
    \left(-\partial^{2}_{z}+\frac{9}{16}\,(k_{\pm})^2\,z^{2}+\frac{3}{2}\,k_{\pm}+\frac{15}{4}\,\frac{1}{z^{2}}\right) f^{n}_{L/R}(z)=M^{2}_{n}\, f^{n}_{L/R}(z), \label{ModSWM}
\end{equation}
for the left $(+)$ and right $(-)$ modes.
These equations are also obtained when the relations \eqref{m5metilde3/2} are substituted into the effective Schr\"odinger equations for the deformed AdS model, Eq. \eqref{AppEq}, with $z$ replaced by $\tilde z$, for both the left and right modes.

The term proportional to $(k_{\pm})^2$ is the same for the left and right modes, together with the centrifugal barrier, assuring a potential well bounded from below in both cases. However, the term proportional to  $k_{\pm}$ have opposite signs implying a shift in the energy between the left and right modes.

The spectrum for the left modes, obtained from Eq. \eqref{ModSWM} with $m_{5+}=\Tilde m_{5+}=3/2$, and  $k_{+}=(4/3)\,\kappa^{2}$, 
in both models (from Eqs. \eqref{MSWM}, and \eqref{DefEspQuad+}, upper signs), is the following: 
\begin{equation}\label{spleft}
     M_{nL}^{2}= 3 k_{+}\, (n+2)= 4 \kappa^2 (n+2);\qquad \qquad  (n=0, 1, 2, \dots). 
\end{equation}

On the other hand, the spectrum for the right modes, also obtained from Eq. \eqref{ModSWM}, now with $m_{5-}=\Tilde m_{5-}=- 3/2$ and $k_{-}=-(4/3)\,\kappa^{2}$, in both models (from Eqs. \eqref{MSWM12},  and  \eqref{DefEspQuad-}, lower signs), is: 
\begin{equation}\label{spright}
     M_{nR}^{2}= 3 |k_{-}|\, (n+1)= 4 \kappa^2 (n+1);\qquad \qquad  (n=0, 1, 2, \dots). 
\end{equation}
Note that the spectrum for the left and right modes, Eqs. \eqref{spleft} and \eqref{spright}, respectively, are different due to the energy shift caused by the term 
$(3/2)k_{\pm}$. 
Actually, they are {\sl almost} degenerate,   except for the ground states. 

So, clearly, the above spectra analysis shows that the dressed soft wall and the  deformed AdS models for  fermions, despite some similarities,  are {\sl not} equivalent, even in the quadratic approximation of the latter.


\subsection{Wave Functions}

In order to obtain a more complete picture in the comparison of the dressed soft wall and the quadratic approximation for the deformed AdS models, here we obtain the analytic wave functions for both. 

As shown in the previous section, for  $m_{5\pm}=\Tilde m_{5\pm}=\pm 3/2$, one finds the one dimensional equations of motion in both models in terms of the holographic coordinate $\zeta$
\begin{equation}
    \left(-\partial^{2}_{\zeta}+\frac{9}{16}\,(k_{\pm})^{2}\,\zeta^{2} + \frac{3}{2}\,k_{\pm}+\frac{15}{4}\,\frac{1}{\zeta^{2}}\right) f^{n}_{L/R}(\zeta)=M^{2}_{n}\, f^{n}_{L/R}(\zeta), \label{SWzeta}
\end{equation}
for the left and right modes, respectively, or equivalently
\begin{equation}
    \left(-\partial^{2}_{\zeta}+\,\kappa^{4}\,\zeta^{2}\pm 2\,\kappa^2+\frac{15}{4}\,\frac{1}{\zeta^{2}}\right) f^{n}_{L/R}(\zeta)=M^{2}_{n}\, f^{n}_{L/R}(\zeta),\label{SWzeta2}
\end{equation}
where $\zeta$ stands for $z$ in the dressed soft wall and for $\tilde z$ in the deformed AdS models. 
 Then, one finds that the normalizable solutions for the left and right modes are given exactly by the same expression in terms of $\zeta$ in both models
\begin{subequations}
\begin{align}
    {f}^{n}_{_{L/R}}(\zeta)&=\sqrt{\frac{2\,\Gamma(n+1)}{\Gamma(n+3)}}\,|\kappa|^{3}\,\zeta^{5/2}\,e^{-\kappa^{2}\zeta^{2}/2}\,L^{2}_{n}(\kappa^{2}\zeta^{2}), 
\end{align}
\end{subequations}
for $n=0, 1, 2, \dots$, since the terms $\pm 2\,\kappa^2$ only produce a shift in the energy of the modes. 

So, the wave functions obtained for the dressed soft wall and the quadratic approximation for the deformed AdS have the same dependence on $\zeta$. However, since $\zeta$ is identified with $z$ in the dressed soft wall and as $\tilde z$ in the deformed AdS model, actually these solutions are not equivalent. In the UV region in both models ($\zeta\to 0$) this difference vanishes, but in the IR ($\zeta\to \infty$) this difference is relevant.


\subsection{Comments on the DIS in both models}

In this section, we briefly comment on the DIS obtainable from the dressed soft wall and the quadratic approximation of the deformed AdS model. The holographic approach to DIS starts with an interaction action which is an integral over the virtual photonic and the baryonic initial and final states. From this action it is possible  to obtain, for instance, the structure functions $F_1(x,q)$ and $F_2(x,q)$, where $x$ is the momentum fraction of the parton, also called the Bjorken variable, defined as $x=q^2/2p\cdot q$, with $q^\mu$ and $p^\mu$ the momenta of the virtual photon and of the initial baryon, respectively \cite{Polchinski:2002jw, BallonBayona:2007qr, Braga:2011wa, FolcoCapossoli:2020pks}. 

The relevant interaction action for DIS in the dressed soft wall model is given by \cite{BallonBayona:2007qr, Braga:2011wa} 
\begin{equation}\label{DISActionDSW}
    S_{int}= g_V \int dz\,d^{4}x\,\sqrt{-g}\,e^{-\kappa^2 z^2}\, \frac{z}{\Omega}\, A_m\bar\Psi_X\,\gamma^{m}\,\Psi_i,
\end{equation}
where $g_V$ is the coupling constant of the interaction of the electromagnetic field $A_m=A_m(x,z)$ with the initial $\Psi_i=\Psi_i(x,z)$ and final $\Psi_X=\Psi_X(x,z)$ fermion states. The fermionic states are given by Eq. \eqref{Psi}, while the electromagnetic field is written in terms of the confluent hypergeometric function of the second kind \cite{BallonBayona:2007qr, Braga:2011wa}: 
\begin{subequations}
    \begin{eqnarray}\label{A_mu}
        A_\mu(x^\mu,z) 
    &=& \eta_\mu e^{i q\cdot x} \kappa^2 z^2\, \Gamma\left( 1 + \frac{q^2}{4\kappa^2}
    \right)  \, {\cal U}\left(
    1 + \frac{q^2}{4\kappa^2}; 2; \kappa^2 z^2 
    \right),\\
     A_z(x^\mu,z) 
    &=& \frac i 2 \eta\cdot q \, z\, e^{i q\cdot x} \, \Gamma\left( 1 + \frac{q^2}{4\kappa^2}
    \right) \,\, {\cal U}\left(
    1 + \frac{q^2}{4\kappa^2}; 1; \kappa^2 z^2 
    \right), \label{A_z}
    \end{eqnarray}
\end{subequations}
where $q_\mu$ and $\eta_\mu$ are the momentum and polarization of the virtual photon, respectively.  

On the other side, the DIS interaction action in the deformed AdS model is given by \cite{FolcoCapossoli:2020pks}
\begin{equation}\label{DISActionDAdS}
    {\tilde S}_{int}= \tilde g_V \int d\tilde z\,d^{4}x\,\sqrt{-\tilde g}\, \frac{\tilde z}{\tilde\Omega}\, \tilde A_m\tilde{\bar\Psi}_X\,\gamma^{m}\,\tilde{\Psi}_i,
\end{equation}
where the fermionic fields are given by Eq. \eqref{Pcompts}, whilst the gauge fields are  described by 
\begin{subequations}
    \begin{eqnarray}\label{A_mutilde}
        \tilde{A}_\mu(x^\mu,\tilde z) 
    &=& - \eta_\mu e^{i q\cdot x} \frac{k}{2}  \tilde z^2 \, \Gamma\left( 1 - \frac{q^2}{2k}
    \right)  \, {\cal U}\left(
    1 - \frac{q^2}{2k}; 2; - \frac{k}{2} \tilde z^2 
    \right),\\
     \tilde{A}_{\tilde z}(x^\mu,\tilde z) 
    &=& \frac i 2 \eta\cdot q \, \tilde z\, e^{i q\cdot x} \, \Gamma\left( 1 - \frac{q^2}{2k}
    \right) \,\, {\cal U}\left(
    1 - \frac{q^2}{2k}; 1; - \frac{k}{2} \tilde z^2 
    \right). 
    \end{eqnarray}
\end{subequations}
Since $k$ is considered negative for the photon fields in this model (see \cite{MartinContreras:2021yfz} for details), these solutions have the same dependence on the holographic coordinate $\zeta$ (introduced in Eqs. \eqref{SWzeta} and \eqref{SWzeta2}) as those presented for the dressed soft wall model, Eqs. \eqref{A_mu} and \eqref{A_z}. This is also the case of the fermionic fields discussed in the previous section.  

Further, we see that there are additional differences in the description of the DIS in the interaction actions, Eqs.  \eqref{DISActionDSW} and \eqref{DISActionDAdS}. Clearly, the presence of the dilaton in action of the dressed soft wall and the warp factor in the deformed AdS model bring in some distictions between them, although the structure functions that can be obtained from these interaction actions in the dressed soft wall and the (full) deformed AdS models are comparable, as can be seen in Refs. \cite{Braga:2011wa}
and \cite{FolcoCapossoli:2020pks}.


\section{Conclusions}

In this work, we have studied the dressed soft wall and the deformed AdS models for fermions and compared their effective Schr\"odinger equations and spectra. This comparison was motivated by recent results on proton DIS structure functions obtained from the deformed AdS model \cite{FolcoCapossoli:2020pks} which resemble that of the dressed soft wall model \cite{Braga:2011wa}. These two models have in common a quadratic exponential factor, which in the case of the dressed soft wall is included in the action and in the mass term of the AdS bulk fermions. On the other hand, the deformed AdS model contains this exponential factor just in the metric which  is asymptotically AdS in the UV. In particular for fermions, the dressed soft wall model is analytically solvable, while the deformed AdS model needs numerical solutions. 

In Section II, we studied the dressed soft wall model for fermions and obtained its effective Schr\"odinger equations for the left and right modes and their solutions in terms of associated Laguerre functions. Then, we obtained the corresponding fermionic spectra including some modifications as the hyperfine spin-spin and meson cloud interactions as in Ref. \cite{Gutsche:2011vb}, as well as anomalous dimensions as considered in Ref. \cite{FolcoCapossoli:2020pks}. 

In Section III, we discussed the deformed AdS model for fermions and obtained its effective Schr\"odinger equations for the left and right modes for which numerical solutions were found in 
\cite{FolcoCapossoli:2019imm, FolcoCapossoli:2020pks}. So, in order to make a comparison with the dressed soft wall model, we perform a power expansion of the effective exponential of the deformed model and consider a quadratic approximation which leads to analytical solutions given by the associated Laguerre functions with corresponding spectra. In Section IV, we comment on the differences between the choices of the sign of the dilaton in both models. In Section V, we present a numerical comparison of the various spectra associated with the dressed soft wall model, discussed in Section II, and the numerical spectra for the deformed AdS model for $N(1/2^+)$ baryons. 

In Section VI, we present a comparison between the dressed soft wall model and the quadratic approximation for the deformed AdS model, discussed in Sections II and III. These two models are analytically solvable for fermions with solutions given by the associated Laguerre functions, although they are not exactly equivalent. Actually, we find relations between these two models and obtain the necessary conditions to relate the left and right modes for each case. We find that when $m_5=\tilde m_5$, we discover that $m_5=\tilde m_5 =\pm 3/2$, where the upper (lower) sign refers to the left (right) modes. This means that to fully describe the left and right modes in the dressed soft wall model from the quadratic approximation of the deformed model one needs to consider fermionic bulk masses with opposite signs. 
It is interesting to note that the AdS/CFT correspondence implies a relation between the fermionic bulk dimensionless parameter $m_5$ and the conformal dimension $\Delta$ of the dual boundary operator: $|m_5|=\Delta -2$, so that the left and right modes discussed here with different signs for $m_5$ and $\tilde m_5$ actually are dual to operators with the same conformal dimension $\Delta$. It is also important to mention that the parameters $m_5$ and $\tilde m_5$ are not the fermionic bulk massses, but are related to these masses as $m_5=\mu_5\Omega$ and $\tilde m_5=\tilde\mu_5\tilde\Omega$, where $\Omega$ and $\tilde\Omega$ are the AdS radii in the two models, which remain arbitrary. So, in some sense, one has some  freedom in establishing the relations between the two models. We have also shown that the analytical wave functions of both models have the same mathematical description in terms of the holographic coordinate $zeta$, which is identified with $z$ for the dressed soft wall and $\tilde z$ for the deformed AdS model. Finally, we comment on the DIS description of both models and show that they have many similarities but are {\sl not} mathematically equivalent. 

The study presented here could help understand better structure functions and form factors, as discussed, for instance,  in Refs. \cite{Abidin:2009hr, Mamo:2021cle, Mamo:2021jhj, Braga:2011wa, FolcoCapossoli:2020pks, Contreras:2021epz}. This is currently under investigation.

\section*{Acknowledgments}

We would like to thank Kostas Rigatos for discussions and an anonymous referee for suggestions.  A.C.P.N. is supported by Coordenação de Aperfeiçoamento de Pessoal de Nível Superior (CAPES). H.B.-F. is  partially supported by Conselho Nacional de Desenvolvimento Cient\'{\i}fico e Tecnol\'{o}gico (CNPq) under grant $\#$ 311079/2019-9.


\appendix


\section{The isotonic oscillator}
\label{isotonic}

In this Appendix, we collect the basic results needed in this work from the quantum mechanical isotonic oscillator, described by the Schrödinger equation (see, {\sl e. g.}, \cite{Ikhdair})
\begin{equation}
    \left(-\partial^{2}_{z}-\epsilon_n +\,\beta^2\,z^{2} +\frac{\alpha}{z^{2}}\right) \psi_{n}(z)=0, \label{IsotonicDE}
\end{equation}
which for $z>0$, $\beta>0$, and $\alpha>- \frac 14$ gives the spectrum 
\begin{equation}
    \epsilon_n = 2\,\beta\, \left(2n + 1 + \xi\right); \qquad (n=0, 1, 2, \dots),  \label{IsoSpec}
\end{equation}
where $\xi=\frac 12 \sqrt{1+4\alpha}>0$, and imply the normalized wave functions
\begin{align}
    \psi_n(z)=\sqrt{\frac{2\,\beta^{\xi+1}n!}{\Gamma(n+\xi+1)}}\,z^{\xi+1/2}\,e^{-\beta z^{2}/2}\,L^{\xi}_{n}(\beta z^{2}). 
\end{align}
%


\section{Case $(ii)$:  $m_5=-\Tilde m_{5} \mp 1$}\label{case(ii)}

In this Appendix, we analyse the case  $m_5=-\Tilde m_{5} \mp 1$, Eq. \eqref{1}. 
Inserting this relation in Eq. \eqref{2}, one gets 
\begin{eqnarray}
2\,\kappa^{2}\,\left(-\Tilde m_{5}\mp \frac{3}{2}\right)=\Tilde m_{5}\,\left(\Tilde m_{5}\mp \frac{1}{2}\right)\,k.\label{sndSoleqn2K}
\end{eqnarray}
Then, squaring this equation and substituting in it the ratio $\kappa^4/k^2$ from Eq. (\ref{3}), one obtains the cubic equation 
\begin{equation}
    (\Tilde m_{5})^3 \pm  \frac{11}{2}(\Tilde m_{5})^2 - \frac{1}{4}\Tilde m_{5} \mp  \frac{27}{8}=0, 
\end{equation}
which has real roots 
\begin{equation}\label{roots}
    \tilde m_{5\frac{L}{R}}^{(1)}=\mp 5.43163; \qquad
    \tilde m_{5\frac{L}{R}}^{(2)}=\mp 0.82319; \qquad 
    \tilde m_{5\frac{L}{R}}^{(3)}=\pm 0.75482,
\end{equation}
with the corresponding ratios 
\begin{equation}\label{kappa/k}
   \left(\frac{\kappa^2}{k}\right)_{\frac{L}{R}}^{(1)}=\pm 4.09734;  \qquad
   \left(\frac{\kappa^2}{k}\right)_{\frac{L}{R}}^{(2)}=\mp 0.804684; \qquad 
   \left(\frac{\kappa^2}{k}\right)_{\frac{L}{R}}^{(3)}=\mp 0.0426516.
\end{equation}
This way, from the effective Schr\"odinger equations for the dressed soft wall, Eq. \eqref{SW}, we obtain that these solutions satisfy
\begin{equation}
    \left[-\partial_{z}^{2}+\kappa^{4}\,z^{2}-2\,\kappa^{2}\left(\tilde m_5\pm \frac{3}{2} \right)+\frac{\tilde m_5(\tilde m_5\pm 1)}{z^{2}} \right] {f}^{n}_{L/R}(z)=M_{n}^{2}\, {f}^{n}_{L/R}(z),\label{SWii}
\end{equation}
with mass spectrum
\begin{equation}
    M_{nL/R}^2=  
     \kappa^2 \left[4n+2 + \left|2\tilde m_5 +1 \right|
    - 2 \tilde m_5 \mp 3\right]; \qquad  (n=0,1,2,\dots).
\end{equation}
Note that all values of $\tilde m^{(i)}_{L/R}$ given by Eq. \eqref{roots}, satisfy the stability condition of the normalizable solutions, $ \tilde m_5(\tilde m_5\pm 1)>-1/4$ (see Appendix \ref{isotonic}).

On the other side, the effective Schr\"odinger equations coming from the deformed AdS model, Eq. \eqref{AppEq}, read 
\begin{equation}
    \left[-\partial^{2}_{z}+\Tilde m_{5}\left(\frac{\Tilde m_{5}}{2}\mp\frac{3}{8}\right)\,k^{2}\,z^{2}+\Tilde m_{5}\left(\Tilde m_{5}\mp\frac{1}{2}\right)\,k+\frac{\Tilde m_{5}\,(\Tilde m_{5}\pm 1)}{z^{2}}\right]\,\psi_{L/R}^{n}(z)=M_{n}^{2}\,\psi_{L/R}^{n}(z).
\end{equation}
which are equivalent to Eq. \eqref{SWii} for the masses given by \eqref{roots} and ratios Eq. \eqref{kappa/k}. The corresponding spectra are given by Eq. \eqref{DefEspQuad+} for $\tilde m_5>1/2$
\begin{subequations}
\begin{align}
    M_{nL/R}^{2>}
        &=
        (\Tilde m_{5}^{2}\mp \frac{\Tilde m_{5}}{2})\,k+2\,|k|\,\sqrt{\left(\frac{\Tilde m_{5}^{2}}{2}\mp \frac{3\,\Tilde m_{5}}{8}\right)}\,\left(2n+1+ \tilde m_5 \pm \frac12\right), 
\end{align}
and Eq. \eqref{DefEspQuad-}, for $\tilde m_5 \le -1/2$ 
\begin{align}
    M_{nL/R}^{2<}
        &=
        (\Tilde m_{5}^{2}\mp \frac{\Tilde m_{5}}{2})\,k+2\,|k|\,\sqrt{\left(\frac{\Tilde m_{5}^{2}}{2}\mp\frac{3\,\Tilde m_{5}}{8}\right)}\,\left(2n+1- \tilde m_5 \mp \frac12\right), 
\end{align} 
\end{subequations}
for $\tilde m_5$ and the ratios $\kappa^2/k$ obeying Eqs. \eqref{roots} and \eqref{kappa/k}, respectively. 

So, clearly, also in the case $m_5=-\Tilde m_{5} \mp 1$, this spectra analysis shows that the dressed soft wall and the  deformed AdS models for  fermions, are {\sl not} equivalent, even in the quadratic approximation of the latter.

\section{Comment on relations with top-down models}

 Fermionic top-down models starting with type II-A or II-B string theory with various configurations with D$p$-branes lead to analytical solutions of the form \cite{Kirsch:2006he,Abt:2019tas,Nakas:2020hyo}
\begin{eqnarray}
    \psi_{\cal G} &=&
     \frac{\varrho^{\ell+1}}{(1+\varrho^2)^{(n+\ell+\frac{m}{2}+\frac{5}{4})}}\, _2F_1\left(-n,-n-\ell-\frac{m+1}{2};\ell+\frac{m+3}{2};-\varrho^2 \right)\chi_+ \cr 
     && + 
     \frac{\varrho^{\ell}}{(1+\varrho^2)^{(n+\ell+\frac{m}{2}+\frac{5}{4})}}\, _2F_1\left(-n,-n-\ell-\frac{m+3}{2};\ell+\frac{m+1}{2};-\varrho^2 \right)\chi_-,
\\
    \psi_{\cal F} &=&
     \frac{\varrho^{\ell}}{(1+\varrho^2)^{(n+\ell+\frac{m}{2}+\frac{1}{4})}}\, _2F_1\left(-n,-n-\ell-\frac{m-1}{2};\ell+\frac{m+1}{2};-\varrho^2 \right)\chi_+ \cr 
     && + 
     \frac{\varrho^{\ell+1}}{(1+\varrho^2)^{(n+\ell+\frac{m}{2}+\frac{1}{4})}}\, _2F_1\left(-n,-n-\ell-\frac{m+1}{2};\ell+\frac{m+3}{2};-\varrho^2 \right)\chi_-,
\end{eqnarray}
which depend on the Gauss hypergeometric function 
\begin{equation}
     _2F_1(a,b;c;z)= \sum_{k=0}^\infty \frac{(a)_k (b)_k}{k!(c)_k} z^k, 
\end{equation}
with spectra
\begin{eqnarray}
    \bar{M}_{\cal G}^2=4\left(n+\ell+\frac {m+1}{2}\right)
    \left( n+\ell+\frac{m+3}{2}\right),   
\\ 
    \bar{M}_{\cal F}^2=4\left(n+\ell+\frac {m-1}{2}\right)
    \left( n+\ell+\frac{m+1}{2}\right).   
\end{eqnarray}

On the other side,  the associate Laguerre functions in terms of which we described the analytical solutions of the fermionic soft walls models discussed in this work can be written  as \cite{Abramowitz}
\begin{equation}
    L_n^{(\alpha)}(x)= \frac{\Gamma(\alpha+n+1)}{n!\Gamma(\alpha+1)}\, _1F_1(-n;\alpha+1;x), 
\end{equation}
in terms of the Kummer confluent hypergeometric function 
\begin{equation}
     _1F_1(a;c;z)= \sum_{k=0}^\infty \frac{(a)_k}{k!(c)_k} z^k, 
\end{equation}
where 
\begin{eqnarray}
    (x)_k &\equiv & \Gamma(x+k)/\Gamma(x)\cr
    &=& x(x+1)\cdots (x+k-1)
\end{eqnarray}
are the Pochhmammer symbols.

Then, in the particular case where  $(b)_k=1$, the two problems are directly related. This happens when $b=0$, which imply
\begin{eqnarray}
    0=n+\ell+\frac{m+1}{2}
    = n+\ell+\frac{m+3}{2};  \qquad ({\cal G}), \\
    0=n+\ell+\frac{m-1}{2}
    = n+\ell+\frac{m+1}{2};  \qquad ({\cal F}),
\end{eqnarray}
meaning massless D$p$-brane modes
\begin{eqnarray}
    \bar M^2_{\cal G}=  
    \bar M^2_{\cal F}=0. 
\end{eqnarray}
So, we see that the dressed soft wall and the quadratic approximation for the deformed AdS models discussed here seem to have fewer degrees of freedom than the D$p$-brane models of Refs.  \cite{Kirsch:2006he,Abt:2019tas,Nakas:2020hyo}. 
 


\begin{thebibliography}{99}



\bibitem{Maldacena:1997re}
J.~M.~Maldacena,
``The Large N limit of superconformal field theories and supergravity,''
Adv. Theor. Math. Phys. \textbf{2}, 231-252 (1998)
doi:10.4310/ATMP.1998.v2.n2.a1
[arXiv:hep-th/9711200 [hep-th]].


\bibitem{Aharony:1999ti}
O.~Aharony, S.~S.~Gubser, J.~M.~Maldacena, H.~Ooguri and Y.~Oz,
``Large N field theories, string theory and gravity,''
Phys. Rept. \textbf{323}, 183-386 (2000)
doi:10.1016/S0370-1573(99)00083-6
[arXiv:hep-th/9905111 [hep-th]].



\bibitem{tHooft:1993dmi}
G.~'t Hooft,
``Dimensional reduction in quantum gravity,''
Conf. Proc. C \textbf{930308}, 284-296 (1993)
[arXiv:gr-qc/9310026 [gr-qc]].


\bibitem{Susskind:1994vu}
L.~Susskind,
``The World as a hologram,''
J. Math. Phys. \textbf{36}, 6377-6396 (1995)
doi:10.1063/1.531249
[arXiv:hep-th/9409089 [hep-th]].


\bibitem{Polchinski:2001tt}
J.~Polchinski and M.~J.~Strassler,
``Hard scattering and gauge / string duality,''
Phys. Rev. Lett. \textbf{88}, 031601 (2002)
doi:10.1103/PhysRevLett.88.031601
[arXiv:hep-th/0109174 [hep-th]].




\bibitem{Boschi-Filho:2002xih}
H.~Boschi-Filho and N.~R.~F.~Braga,
``Gauge / string duality and scalar glueball mass ratios,''
JHEP \textbf{05}, 009 (2003)
doi:10.1088/1126-6708/2003/05/009
[arXiv:hep-th/0212207 [hep-th]].


\bibitem{Boschi-Filho:2002wdj}
H.~Boschi-Filho and N.~R.~F.~Braga,
``QCD / string holographic mapping and glueball mass spectrum,''
Eur. Phys. J. C \textbf{32}, 529-533 (2004)
doi:10.1140/epjc/s2003-01526-4
[arXiv:hep-th/0209080 [hep-th]].

\bibitem{Polchinski:2002jw}
J.~Polchinski and M.~J.~Strassler,
``Deep inelastic scattering and gauge / string duality,''
JHEP \textbf{05}, 012 (2003)
doi:10.1088/1126-6708/2003/05/012
[arXiv:hep-th/0209211 [hep-th]].

\bibitem{Erlich:2005qh}
J.~Erlich, E.~Katz, D.~T.~Son and M.~A.~Stephanov,
``QCD and a holographic model of hadrons,''
Phys. Rev. Lett. \textbf{95}, 261602 (2005)
doi:10.1103/PhysRevLett.95.261602
[arXiv:hep-ph/0501128 [hep-ph]].

\bibitem{deTeramond:2005su}
G.~F.~de Teramond and S.~J.~Brodsky,
``Hadronic spectrum of a holographic dual of QCD,''
Phys. Rev. Lett. \textbf{94}, 201601 (2005)
doi:10.1103/PhysRevLett.94.201601
[arXiv:hep-th/0501022 [hep-th]].


\bibitem{Boschi-Filho:2005xct}
H.~Boschi-Filho, N.~R.~F.~Braga and H.~L.~Carrion,
``Glueball Regge trajectories from gauge/string duality and the Pomeron,''
Phys. Rev. D \textbf{73}, 047901 (2006)
doi:10.1103/PhysRevD.73.047901
[arXiv:hep-th/0507063 [hep-th]].

\bibitem{Karch:2006pv}
A.~Karch, E.~Katz, D.~T.~Son and M.~A.~Stephanov,
``Linear confinement and AdS/QCD,''
Phys. Rev. D \textbf{74}, 015005 (2006)
doi:10.1103/PhysRevD.74.015005
[arXiv:hep-ph/0602229 [hep-ph]].


\bibitem{Karch:2010eg}
A.~Karch, E.~Katz, D.~T.~Son and M.~A.~Stephanov,
``On the sign of the dilaton in the soft wall models,''
JHEP \textbf{04}, 066 (2011)
doi:10.1007/JHEP04(2011)066
[arXiv:1012.4813 [hep-ph]].

\bibitem{Colangelo:2008us}
P.~Colangelo, F.~De Fazio, F.~Giannuzzi, F.~Jugeau and S.~Nicotri,
``Light scalar mesons in the soft-wall model of AdS/QCD,''
Phys. Rev. D \textbf{78}, 055009 (2008)
doi:10.1103/PhysRevD.78.055009
[arXiv:0807.1054 [hep-ph]].

\bibitem{Colangelo:2007pt}
P.~Colangelo, F.~De Fazio, F.~Jugeau and S.~Nicotri,
``On the light glueball spectrum in a holographic description of QCD,''
Phys. Lett. B \textbf{652}, 73-78 (2007)
doi:10.1016/j.physletb.2007.06.072
[arXiv:hep-ph/0703316 [hep-ph]].


\bibitem{Gursoy:2007cb}
U.~Gursoy and E.~Kiritsis,
``Exploring improved holographic theories for QCD: Part I,''
JHEP \textbf{02}, 032 (2008)
doi:10.1088/1126-6708/2008/02/032
[arXiv:0707.1324 [hep-th]].


\bibitem{Gursoy:2007er}
U.~Gursoy, E.~Kiritsis and F.~Nitti,
``Exploring improved holographic theories for QCD: Part II,''
JHEP \textbf{02}, 019 (2008)
doi:10.1088/1126-6708/2008/02/019
[arXiv:0707.1349 [hep-th]].


\bibitem{Li:2013oda}
D.~Li and M.~Huang,
``Dynamical holographic QCD model for glueball and light meson spectra,''
JHEP \textbf{11}, 088 (2013)
doi:10.1007/JHEP11(2013)088
[arXiv:1303.6929 [hep-ph]].



\bibitem{BallonBayona:2007qr}
C.~A.~Ballon Bayona, H.~Boschi-Filho and N.~R.~F.~Braga,
``Deep inelastic scattering from gauge string duality in the soft wall model,''
JHEP \textbf{03}, 064 (2008)
doi:10.1088/1126-6708/2008/03/064
[arXiv:0711.0221 [hep-th]].


\bibitem{Abidin:2009hr}
Z.~Abidin and C.~E.~Carlson,
``Nucleon electromagnetic and gravitational form factors from holography,''
Phys. Rev. D \textbf{79}, 115003 (2009)
doi:10.1103/PhysRevD.79.115003
[arXiv:0903.4818 [hep-ph]].








\bibitem{Branz:2010ub}
T.~Branz, T.~Gutsche, V.~E.~Lyubovitskij, I.~Schmidt and A.~Vega,
``Light and heavy mesons in a soft-wall holographic approach,''
Phys. Rev. D \textbf{82}, 074022 (2010)
doi:10.1103/PhysRevD.82.074022
[arXiv:1008.0268 [hep-ph]].


\bibitem{Gutsche:2011vb}
T.~Gutsche, V.~E.~Lyubovitskij, I.~Schmidt and A.~Vega,
``Dilaton in a soft-wall holographic approach to mesons and baryons,''
Phys. Rev. D \textbf{85} (2012), 076003
doi:10.1103/PhysRevD.85.076003
[arXiv:1108.0346 [hep-ph]].





%
\bibitem{Braga:2011wa}
N.~R.~F.~Braga and A.~Vega,
``Deep inelastic scattering of baryons in a modified soft wall model,''
Eur. Phys. J. C \textbf{72} (2012), 2236
doi:10.1140/epjc/s10052-012-2236-2
[arXiv:1110.2548 [hep-ph]].


\bibitem{FolcoCapossoli:2019imm}
E.~Folco Capossoli, M.~A.~Mart\'\i{}n Contreras, D.~Li, A.~Vega and H.~Boschi-Filho,
``Hadronic spectra from deformed AdS backgrounds,''
Chin. Phys. C \textbf{44}, no.6, 064104 (2020)
doi:10.1088/1674-1137/44/6/064104
[arXiv:1903.06269 [hep-ph]].


\bibitem{FolcoCapossoli:2020pks}
E.~Folco Capossoli, M.~A.~Mart\'\i{}n Contreras, D.~Li, A.~Vega and H.~Boschi-Filho,
``Proton structure functions from an AdS/QCD model with a deformed background,''
Phys. Rev. D \textbf{102}, no.8, 086004 (2020)
doi:10.1103/PhysRevD.102.086004
[arXiv:2007.09283 [hep-ph]].


%
\bibitem{MartinContreras:2021yfz}
M.~A.~Martin Contreras, E.~Folco Capossoli, D.~Li, A.~Vega and H.~Boschi-Filho,
``Pion form factor from an AdS deformed background,''
Nucl. Phys. B \textbf{977}, 115726 (2022)
doi:10.1016/j.nuclphysb.2022.115726
[arXiv:2104.04640 [hep-ph]].


%
\bibitem{Contreras:2021epz}
M.~A.~M.~Contreras, E.~F.~Capossoli, D.~Li, A.~Vega and H.~Boschi-Filho,
``Proton and neutron form factors from deformed gravity/gauge duality,''
Phys. Lett. B \textbf{822}, 136638 (2021)
doi:10.1016/j.physletb.2021.136638
[arXiv:2108.05427 [hep-ph]].


\bibitem{Caldeira:2020sot}
N.~G.~Caldeira, E.~Folco Capossoli, C.~A.~D.~Zarro and H.~Boschi-Filho,
``Fluctuation and dissipation from a deformed string/gauge duality model,''
Phys. Rev. D \textbf{102}, no.8, 086005 (2020)
doi:10.1103/PhysRevD.102.086005
[arXiv:2007.00160 [hep-th]].

%
\bibitem{Caldeira:2020rir}
N.~G.~Caldeira, E.~Folco Capossoli, C.~A.~D.~Zarro and H.~Boschi-Filho,
``Fluctuation and dissipation within a deformed holographic model with backreaction,''
Phys. Lett. B \textbf{815}, 136140 (2021)
doi:10.1016/j.physletb.2021.136140
[arXiv:2010.15293 [hep-th]].

\bibitem{Caldeira:2021izy}
N.~G.~Caldeira, E.~F.~Capossoli, C.~A.~D.~Zarro and H.~Boschi-Filho,
``Fermionic and bosonic fluctuation-dissipation theorem from a deformed AdS holographic model at finite temperature and chemical potential,''
Eur. Phys. J. C \textbf{82}, no.1, 16 (2022)
doi:10.1140/epjc/s10052-021-09963-3
[arXiv:2104.08397 [hep-th]].

\bibitem{Andreev:2006ct}
O.~Andreev and V.~I.~Zakharov,
``Heavy-quark potentials and AdS/QCD,''
Phys. Rev. D \textbf{74}, 025023 (2006)
doi:10.1103/PhysRevD.74.025023
[arXiv:hep-ph/0604204 [hep-ph]].


\bibitem{Bruni:2018dqm}
R.~C.~L.~Bruni, E.~Folco Capossoli and H.~Boschi-Filho,
``Quark-antiquark potential from a deformed AdS/QCD,''
Adv. High Energy Phys. \textbf{2019}, 1901659 (2019)
doi:10.1155/2019/1901659
[arXiv:1806.05720 [hep-th]].

\bibitem{Boschi-Filho:2012ijd}
H.~Boschi-Filho, N.~R.~F.~Braga, F.~Jugeau and M.~A.~C.~Torres,
``Anomalous dimensions and scalar glueball spectroscopy in AdS/QCD,''
Eur. Phys. J. C \textbf{73}, 2540 (2013)
doi:10.1140/epjc/s10052-013-2540-5
[arXiv:1208.2291 [hep-th]].


\bibitem{FolcoCapossoli:2016uns}
E.~Folco Capossoli, D.~Li and H.~Boschi-Filho,
``Dynamical corrections to the anomalous holographic soft-wall model: the pomeron and the odderon,''
Eur. Phys. J. C \textbf{76}, no.6, 320 (2016)
doi:10.1140/epjc/s10052-016-4171-0
[arXiv:1604.01647 [hep-ph]].


\bibitem{Rodrigues:2016kez}
D.~M.~Rodrigues, E.~Folco Capossoli and H.~Boschi-Filho,
``Scalar and higher even spin glueball masses from an anomalous modified holographic model,''
EPL \textbf{122}, no.2, 21001 (2018)
doi:10.1209/0295-5075/122/21001
[arXiv:1611.09817 [hep-ph]].


\bibitem{Kirsch:2006he}
I.~Kirsch,
``Spectroscopy of fermionic operators in AdS/CFT,''
JHEP \textbf{09}, 052 (2006)
doi:10.1088/1126-6708/2006/09/052
[arXiv:hep-th/0607205 [hep-th]].

\bibitem{Abt:2019tas}
R.~Abt, J.~Erdmenger, N.~Evans and K.~S.~Rigatos,
``Light composite fermions from holography,''
JHEP \textbf{11}, 160 (2019)
doi:10.1007/JHEP11(2019)160
[arXiv:1907.09489 [hep-th]].

\bibitem{Nakas:2020hyo}
T.~Nakas and K.~S.~Rigatos,
``Fermions and baryons as open-string states from brane junctions,''
JHEP \textbf{12}, 157 (2020)
doi:10.1007/JHEP12(2020)157
[arXiv:2010.00025 [hep-th]].






\bibitem{Workman:2022ynf}
R.~L.~Workman \textit{et al.} [Particle Data Group],
``Review of Particle Physics,''
PTEP \textbf{2022}, 083C01 (2022)
doi:10.1093/ptep/ptac097


\bibitem{Mamo:2021cle}
K.~A.~Mamo and I.~Zahed,
``Neutrino-nucleon DIS from holographic QCD: PDFs of sea and valence quarks, form factors, and structure functions of the proton,''
Phys. Rev. D \textbf{104} (2021) no.6, 066010
doi:10.1103/PhysRevD.104.066010
[arXiv:2102.00608 [hep-ph]].

\bibitem{Mamo:2021jhj}
K.~A.~Mamo and I.~Zahed,
``Electromagnetic radii of the nucleon in soft-wall holographic QCD,''
[arXiv:2106.00752 [hep-ph]].


\bibitem{Ikhdair}
Ikhdair, S. M, Sever, R. 
``Relativistic and nonrelativistic bound states of the isotonic oscillator by Nikiforov-Uvarov method,'' 
J.  Math. Phys. \textbf{52}, 122108 (2011); 
doi:10.1063/1.3671640 
[arXiv:1203.1736].



\bibitem{Abramowitz} 
Abramowitz, M. and Stegun, I. A. (Eds.). Handbook of Mathematical Functions with Formulas, Graphs, and Mathematical Tables, 9th printing. New York: Dover, 1972.








\end{thebibliography}
\end{document}